\documentclass[accepted]{article}

\usepackage{microtype}
\usepackage{graphicx}
\usepackage{subcaption}
\usepackage{booktabs} 

\usepackage{hyperref}

\usepackage{icml2026}

\usepackage{amsmath}
\usepackage{amssymb}
\usepackage{mathtools}
\usepackage{amsthm}
\usepackage{soul}

\usepackage[capitalize,noabbrev]{cleveref}

\theoremstyle{plain}

\theoremstyle{definition}

\theoremstyle{remark}

\usepackage[textsize=tiny]{todonotes}

\icmltitlerunning{Re-identification of Participants in the Anthropic Interviewer Dataset}

\begin{document}

\twocolumn[
  \icmltitle{Agentic LLMs as Powerful Deanonymizers: Re-identification of Participants in the Anthropic Interviewer Dataset}



  \icmlsetsymbol{equal}{*}

  \begin{icmlauthorlist}
    \icmlauthor{Tianshi Li}{yyy}
  \end{icmlauthorlist}

  \icmlaffiliation{yyy}{Khoury College of Computer Sciences, Northeastern University, Boston, MA, USA}

  \icmlcorrespondingauthor{Tianshi Li}{tia.li@northeastern.edu}


  \vskip 0.3in
]



\printAffiliationsAndNotice{}  

\begin{abstract}
On December 4, 2025, Anthropic released Anthropic Interviewer, an AI tool for running qualitative interviews at scale, along with a public dataset of 1,250 interviews with professionals, including 125 scientists, about their use of AI for research. Focusing on the scientist subset, I show that widely available LLMs with web search and agentic capabilities can link six out of twenty-four interviews to specific scientific works, recovering associated authors and, in some cases, uniquely identifying the interviewees.
My contribution is to show that modern LLM-based agents make such re-identification attacks easy and low-effort: off-the-shelf tools can, with a few natural-language prompts, search the web, cross-reference details, and propose likely matches, effectively lowering the technical barrier. Existing safeguards can be bypassed by breaking down the re-identification into benign tasks.
I outline the attack at a high level, discuss implications for releasing rich qualitative data in the age of LLM agents, and propose mitigation recommendations and open problems. I have notified Anthropic of my findings.

\end{abstract}

\section{Introduction and Background}

On December 4, 2025, Anthropic introduced Anthropic Interviewer, an AI-powered tool for conducting qualitative interviews at scale~\citep{handa2025interviewer}.
As part of this launch, Anthropic conducted a large-scale interview study with 1{,}250 professionals — the general workforce (N=1{,}000), scientists (N=125), and creatives (N=125) — and publicly released all interview transcripts on Hugging Face~\citep{anthropicInterviewerDataset2025}.

The dataset release is justified by participant consent.
However, the full consent process is not disclosed, and
participants may reasonably be under the impression that the data release is anonymized: for example, the interview script of the general workforce\footnotemark subset reassures them that \textit{``I'll be taking notes during our chat, but rest assured that anything you share won't be personally attributed to you. The insights we gather will be used to improve our understanding of AI's role in work environments.''} --- emphasizing the benefits yet downplaying the re-identification risks.
\footnotetext{The general workforce and scientists are two different subsets of the dataset released by Anthropic.}


I focus on the scientist subset of the dataset and demonstrate a practical re-identification attack powered by large language models (LLMs) augmented with web search.
By matching project details described in interview transcripts with publicly available publications, my method can often recover the underlying paper and, correspondingly, narrow down the scope or even uniquely identify the interviewee.
Anthropic has taken steps to anonymize the data (e.g., redacting certain details), yet my experiments show that at least some of the redacted information can be easily recovered.

My attack highlights the realistic privacy risks created when rich qualitative data is released, as powerful, general-purpose LLM tools have become widely available, whose use is difficult to constrain.
In particular, I show that LLM safeguards can be bypassed by breaking down the attack into benign tasks, exploiting both the dual-use nature of information-retrieval tools and the inherent unverifiability of user intent.
\textbf{The barrier to carrying out this attack is extremely low: anyone with access to an LLM agent with web search can use a small number of natural-language prompts to robustly perform the attack.}
I chose to disclose the methods at a high level to illustrate the general risks without providing details to facilitate harm to the participants.
\textbf{The main purpose of this paper is to document this case and serve as a timely reminder to the research community and the general public about this serious, pervasive vulnerability, so we can collectively consider its implications and potential mitigations.
}

\section{Method}

\paragraph{Re-identification attack setting} My attack operates on the scientist subset of the dataset. Each interview is initiated by the AI interviewer asking the participant to ``walk me through a recent research project you’ve worked on,'' including how the project evolved from ``initial idea'' to ``final output,''  eliciting rich information about the research project. As auxiliary data, I assume access to arbitrary information on the public internet, operationalized through web-augmented LLM services that can search, retrieve, and summarize relevant publications.

\paragraph{Intuition} My attack relies on two observations. First, research projects are often described using domain-specific terminology, niche problem settings, and distinctive contributions, making them technically effective quasi-identifiers; Second, web-augmented LLMs can dramatically accelerate the search, synthesis, matching, and ranking required for re-identification, \textbf{turning what would otherwise be a time-consuming manual process that require both technical and domain expertise into one that can be executed with a handful of natural language prompts and in matters of minutes, by anyone.}

\paragraph{Procedure}
My re-identification procedure consists of two main steps. First, for each full interview transcript in the scientist subset, I use a non-thinking model to label whether the interviewee discusses specific published work (e.g., a paper, dissertation, etc.) and assign a numerical label indicating how many distinct published projects are mentioned (0 if none is mentioned). This yields a filtered subset of interviews whose project descriptions indicate at least one published work. Second, for each of these interviews, I call a thinking model agent to search for candidate publications that match the described project and return a ranked list (web search enabled). For every candidate, the system records a discrete confidence rating --- ``very low,'' ``low,'' ``medium,'' ``high,'' or ``very high'' --- together with two short rationales explaining how the publication does and does not align with the project described in the transcript. I run this process multiple times, appending to the candidate sets until no new ``very high'' confidence matches are discovered.
I use ``very high'' as a threshold for potential re-identification.
To mitigate the risk of exposing re-identified participants, I intentionally omit operational details on how to scale the procedure or circumvent LLM safeguards.

\section{Results}

Out of the 125 scientist interview transcripts, 24 were detected as mentioning at least one publication, and then re-identification attempts were performed with them. Among these 24, I was able to recover the specific publication(s) being discussed for 6 transcripts (25\%). Notably, some of these cases involved a thesis or dissertation, which is single-authored and therefore uniquely identifies the interviewee.

\paragraph{Confidence} Seven cases ended up receiving a ``very high'' label by the LLM. 
I manually verified all of them and determined that the re-identifications for six interview transcripts are indeed of very high confidence.
These interview descriptions matched multiple aspects of the corresponding publications, such as methodology, procedures, key contributions, outcomes, timelines, and team compositions.
These matches involved both specific keywords and semantically aligned descriptions expressed in richer natural language that do not map cleanly to simple categories or verbatim phrases from the paper.
In several instances, participants did not reveal all details at once; instead, the AI interviewer’s follow-up questions elicited additional information that, in aggregate, made the project highly identifiable.
During manual verification, I reviewed both the full interview transcripts and the original publications to confirm that highly granular and non-trivial details were consistent, and that the overlap in technical language and jargon was substantial.

In the one case where the LLM assigned very high confidence but I did not count it as a success, both the interviewee’s description and my search results indicate that the main project they described is likely still in the review phase.
However, my method still identified papers with a highly overlapping set of authors, which demonstrated a trajectory of research methodologies and contributions closely matching the interview descriptions. I did not count this as a success, but I note that re-identification is still probable, especially once the exact paper has been published.

\section{Implications}

\paragraph{Re-identification attacks made easy and scalable}
My experiments programmatically invoked LLM APIs to perform the attacks.
The API cost is low, with each transcript re-identification attempt costing less than \$0.5 with about 4 minutes of run time.
\textbf{I also evaluated a no-code variant of the attack by directly prompting user-facing LLM services that provide web search capabilities, which was similarly successful.
}
Prior research has demonstrated that LLMs plus agentic capabilities can facilitate the aggregation of public information for privacy-invasive tasks such as building user profiles and customizing phishing emails~\cite{mireshghallah2025position,kim2025llms}, and this work provides a concrete example of using them for re-identification attacks that can yield material harms (discussed below).

\paragraph{Harms to participants} Participants in this dataset may experience the following harms when being re-identified.

\textit{Unexpected exposure.}
Participants may not expect their identities to be exposed when giving consent. Thwarted expectations are one of the main categories of privacy harms~\cite{citron2022privacy}.

\textit{Emotional distress.}
The use of AI for work is a sensitive topic that people tend to hide from others and may carry stigma~\cite{10.1145/3711061}. If participants find out that their identities are linked with their self-reported use of AI, it may cause anxiety about how others perceive them.

\textit{Reputational harms.}
Some participants expressed a lack of confidence in their technical skills.
Their methods of AI use may not be considered acceptable, may be viewed as over-reliance on AI, and may invite doubts about the quality of their research and increased scrutiny of specific papers.

\textit{Relationship harms.}
Some participants made comments that implied criticism of or disagreement with their colleagues or collaborators, both regarding research practices and the use of AI. Some participants may also appear to rely on others to avoid errors introduced by AI. If these considerations and practices were known to the people involved, they could harm trust and relationships. In addition, people hold different opinions about AI, and inadvertently revealing one’s opinions about AI can lead to friction in relationships.

\textit{Revelation of policy violations.}
Participants mentioned varied uses of AI, including areas that might be grey zones or subject to policy controls. For example, using AI to perform data analysis may involve sharing data with third-party AI tools in ways that violate institutional policies; using AI to generate writing drafts for papers may violate conference or journal policies. This is more speculative, as the interview transcripts do not provide the legal or procedural details, but some language might invite doubts and scrutiny.

\textit{Risks may increase over time.}
These risks may increase over time because there are papers that are works in progress or under review. More data points may become re-identifiable as additional papers are published. Furthermore, future LLM models and systems may be stronger and have better access to tools and information, such as materials behind paywalls or internal databases. These enhancements to the LLM feature may also make re-identification more feasible.

\section{Responsible Disclosure}

I contacted Anthropic to share my findings one day after the initial release of the tool and dataset.
In my emails, I presented the method, results, and implications of my attack, and attached a few re-identification examples for verification.
I also proposed concrete mitigation suggestions: (1) Take down or at least temporarily hide the dataset; (2) Debrief participants and re-collect consents.

I received responses co-signed by the main contact of the Anthropic Interviewer dataset and the Anthropic privacy team, which reiterated that participants had consented to the public release of the raw interview transcripts and had been advised not to share identifying information if they preferred anonymity.
They considered the redaction as ``an additional courtesy, not a privacy safeguard''.
As a result of the communication, the research team has updated the HF dataset README to clarify that the consent was provided for the public release \textit{of their raw transcripts}.
No further actions have been taken as of January 8, 2026.

\section{Open Problems}

Although being described as a ``disctinct, one-time effort,'' the release of this dataset could set a precedent for future releases of similar datasets. The next phase of the Anthropic Interviewer experiment has already begun data collection. Anthropic’s release page indicates that more people will see ``a pop-up in Claude.ai asking you to participate in interviews''~\cite{handa2025interviewer}. The FAQ further notes that ``We may also include anonymized responses in published findings.'' The problems revealed in this stage of data release need to be recognized and addressed soon to protect future participants from similar risks.

\textbf{Finally, I would like readers to look beyond this single incident and consider the broader challenges facing the research community in responsibly releasing qualitative datasets in the age of LLM agents.
}
Previously, dataset de-anonymization required a lot of manual effort and expertise~\cite{narayanan2006break}.
Qualitative data often includes open-ended details that function as quasi-identifiers, a vulnerability already demonstrated by work linking ChatGPT chat logs with published articles containing machine-generated content~\cite{brigham2024developing}. With emerging agentic capabilities—including tool use, planning, and reasoning at or near human-expert level—this work shows that even high-level descriptions can suffice for linkage, and the time, effort, and expertise required to mount such attacks are drastically reduced.

Many open problems emerge and urgently need to be resolved. Below are some examples.
\begin{itemize}
    \item Since prior redaction practices proved inadequate for anonymization, what privacy guarantees can we realistically provide while balancing privacy and data utility for qualitative data release?
    \item How can we meaningfully inform participants of the risks and ensure they are actually able to remove implicit identifying information if they wish to remain anonymous—or, at minimum, set realistic expectations about re-identification risk?
    \item How can we strengthen model and agent safety guardrails to address the ambiguity of user intent, which can be exploited to circumvent current safeguards by reframing malicious goals as benign tasks?
\end{itemize}
\textbf{I call on the research community to recognize and systematically study these issues before they cause further harm or become normalized.}

\section{Ethics Considerations}

As with any work about a cybersecurity or privacy attack, I consider whether disclosure increases harm by lowering the barrier to abuse, and whether those risks are justified by the expected societal benefits. I assess the marginal risk of publication as limited because (1) LLM agent systems and tooling are broadly accessible, and (2) the attack does not require specialized infrastructure or advanced technical capability, making discovery by motivated attackers plausible and potentially already underway. Conversely, transparent reporting enables the community to recognize the threat, reproduce and measure it under controlled conditions, and develop effective mitigations, detection strategies, and best practices. Accordingly, I conclude that the anticipated benefits of responsible publication outweigh the risks.

I have taken the following steps to further limit risk. I provide enough methodological detail to substantiate the feasibility, scale, confidence, and impact of my findings, while intentionally withholding directly identifiable information and any tools, code, or procedural specifics that would make the attack easy to reproduce. The goal is to communicate the nature of the risk and support mitigation efforts while minimizing potential harm to individuals. Consistent with this approach, I will not publicly release detailed prompts or step-by-step techniques for bypassing safeguards, particularly given the attack’s low barrier to execution (including no-code pathways).

I practiced responsible disclosure, actively reporting the issue to and following up with Anthropic, including Anthropic’s privacy team and researchers on the Societal Impacts team who developed the Anthropic Interviewer tool and maintain the dataset, to address the problem. I first contacted them on December 5, within one day of the project, product, and dataset release.

\section*{Acknowledgements}

I would like to thank Lorrie Cranor, Jason Hong, Tadayoshi Kohno, Alan Mislove, and Jackie Yang, for their valuable feedback on this work.

\bibliography{example_paper}
\bibliographystyle{icml2026}

\newpage
\appendix
\onecolumn



\end{document}